# Crystallization of pristine cubic ice from liquid at ambient pressure

Chenwei Zheng[1], Paul F. Henry[2,3], Aasim I. Shaffi[1], Sanghamitra Mukhopadhyay[2], Miroslava Novoveska[4], Milz L. Beaumont[4], Yidan Wang[1], Camilla Di Mino[5], Tom F. Headen[2], Marta Falkowska[6], Christopher A. Howard[1], Adam J. Clancy[4], Christoph G. Salzmann[4], Neal T. Skipper[1]*

(1) Department of Physics and Astronomy, University College London, London, WC1E 6BT, UK
(2) ISIS Neutron and Muon Source, Science and Technology Facilities Council, Rutherford Appleton Laboratory, Didcot, OX11 0QX, UK
(3) Department of Chemistry, Uppsala Universitet, Box 538, SE-751 21 Uppsala, SWEDEN
(4) Department of Chemistry, University College London, London, WC1H 0AJ, UK
(5) Department of Chemistry, University of Oxford, South Parks Road, Oxford, OX1 3QZ, UK
(6) Department of Chemical Engineering, The University of Manchester, Manchester, M13 9PL, UK

*Correspondence: n.skipper@ucl.ac.uk

## Abstract

The phase diagram of frozen water is famously rich: to date, over twenty crystalline polymorphs have been identified. Of the low-pressure "ice I" family, hexagonal (I*h*) is the principal form on Earth, while cubic (I*c*) is much more elusive. Fundamental questions remain open as to whether cubic ice I*c* can form directly from the liquid state, its thermodynamic stability and natural occurrence. Here we show that pristine cubic ice I*c* can be formed at atmospheric pressure simply by cooling an aqueous solution confined within mesoporous silica. Using primarily neutron scattering, we show unambiguously that under these conditions, cubic ice I*c* forms reproducibly and is the only thermodynamically stable crystalline phase of water. The discovery that cubic ice I*c* is directly accessible from the liquid state, and stable at atmospheric pressure, strongly suggests that this polymorph plays a much more significant role in natural and synthetic processes than previously thought.

## Main

While 21 water ice polymorphs have been identified to date, the overwhelming majority of ice on Earth adopts the hexagonal structure (ice I*h*), which is stable over a wide range of pressures and temperatures[1,2]. Its close relative, cubic ice (ice I*c*), first proposed by König in 1943 from vapour-deposited water[3], arranges tetrahedrally coordinated water molecules into a diamond-cubic structure. Cubic-dominated nuclei are implicated in the earliest stages of homogeneous water nucleation[4], yet the pristine phase itself remains strikingly elusive. Whether pristine cubic ice truly exists in nature has been debated for decades[5-8], though indirect evidence dates back to nearly 400 years ago. For example, a rare optical phenomenon known as Scheiner's halo has historically been considered as evidence for cubic ice I*c* in the Earth's atmosphere[9]. Further observations attributed to cubic ice I*c* include the supersaturation of water vapour in cirrus clouds and the optical behaviour of polar mesospheric clouds[10-12]. Previous studies have also suggested that transient cubic ice I*c* forms within comet nuclei and their dust grains[13-15].

This lack of direct evidence has spurred researchers to investigate various strategies for synthesizing the cubic ice I*c* polymorph, including (i) freezing aqueous solutions[10], (ii) rapid quenching of water droplets[16], (iii) freezing water within confined geometry[17], (iv) thermal treatment of high-pressure ice phases[18,19], (v) warming amorphous ice[20,21], and (vi) dissociating gas hydrates[22-24]. However, it is now recognized that in these attempts, the ice phase in fact contains a mixture of cubic and hexagonal stacking, denoted as stacking-disordered ice (ice I*sd*)[5,6,19]. Fully cubic ice I*c* has only been achieved very recently: del Rosso *et al.*[25] produced pristine cubic ice I*c* by heating $D_2O$ ice XVII powder (obtained from hydrogen-filled $C_0$ clathrate hydrate) under vacuum, Komatsu *et al.*[26] made pristine ice I*c* by decompressing $C_2$ hydrogen hydrate at 100 K, and Huang *et al.*[27] deposited nano crystallites of cubic ice from vapor at ~$10^{-8}$ mbar and 106 K. These methods require complex conditions and processing involving a metastable high-pressure ice precursor and cryogenic decompression, or remarkably low pressures. Critically, all three routes bypass liquid water entirely. Indeed, eight decades after König's proposal, it remains unknown whether and how pristine cubic ice I*c* can form directly from liquid water or aqueous solution, the pathway by which most ice on Earth forms, and remain stable under ambient pressure.

We have established a pathway to pure cubic ice I*c* from aqueous solution at ambient pressure, in an experimental analogue representative of many natural systems. Our strategy is based on clathrate-forming aqueous solution and nanoconfinement. The latter was achieved by using MCM-41, a mesoporous silica characterized by a regular hexagonal array of parallel cylindrical pores (diameter 3.7 nm), whose silanol-rich surface chemistry could mimic the hydrophilic nanocavities found in atmospheric mineral dust[28]. This approach is consistent with a recent simulation, which suggests that such hydroxylated substrates are among the most effective promoters of cubic ice I*c* nucleation[29]. However, previous studies of pure water freezing in such mesoporous silica substrates have indicated the presence of hexagonal and/or stacking disordered ice components in addition to cubic-like features[17, 30-34]. Drawing on the successful isolation of bulk cubic ice I*c* from clathrate hydrate precursors[25,26], we have confined an ambient pressure clathrate-forming liquid in MCM-41, namely: tetrahydrofuran (THF) in water at a molar ratio 1:17. At this composition, bulk aqueous THF forms a structure II clathrate hydrate directly from the liquid at 277.4 K[35]. This hydrate is based on cavity filling of ice XVI and has space group $Fd\bar{3}m$, which is the same space group as cubic ice I*c*.

We have investigated ice formation from the clathrate-forming THF-water liquid mixture (1:17 molar ratio) in MCM-41 as a function of pore filling. A range of different liquid loadings have been studied: (a) 0.34 $cm^3$/g silica; (b) 0.60 $cm^3$/g silica; (c) 0.70 $cm^3$/g silica, and (d) 0.94 $cm^3$/g silica. For reference, the MCM-41 silica pore volume is ~0.78 $cm^3$/g silica (see Methods). The crystallographic details of the solid phases were determined using the high-intensity neutron powder diffractometer (Polaris) at the STFC ISIS Neutron and Muon Source (UK) (see Methods). Deuterated liquids ($C_4D_8O$ and $D_2O$) were used to maximize the diffraction signal from the confined solid, while removing the large incoherent scattering effects associated with hydrogen (H). Diffraction patterns were measured at 150 K and ambient pressure over a broad range of wavevector transfer magnitude, *Q,* with the wide-angle region ($Q \approx 1.0$ - 7.0 $Å^{-1}$; $d \approx$

6.3 - 0.9 Å) used to resolve crystalline ice phases (Fig. 1), and the small-angle region ($Q \approx 1$ – 0.16 Å$^{-1}$, $d \approx 6.3$ - 39.3 Å) to monitor changes associated with the mesoporous host structure (Extended Data Fig. 1).

We note immediately that comparison of the MCM-41 sample with liquid loading 0.34 cm$^3$/g silica versus the empty MCM-41 reveals Bragg peaks at 1.70 Å$^{-1}$, 2.78 Å$^{-1}$, 3.42 Å$^{-1}$ and 4.28 Å$^{-1}$, characteristic of cubic ice I*c*[25,26] (Fig. 1). At a liquid loading of 0.60 cm$^3$/g silica, these peaks become more pronounced, indicative of an increased quantity of this phase, with no signal from other forms of crystalline ice observed. We therefore conclude that pristine cubic ice I*c* forms reproducibly at liquid loadings from 0.34 cm$^3$/g silica to 0.60 cm$^3$/g silica. A separate measurement at the liquid loading of 0.60 cm$^3$/g silica was performed on the OSIRIS spectrometer with diffraction capabilities at 5 K, which was consistent with this finding (Extended Data Fig. 2). Differential Scanning Calorimetry (DSC, see Methods) confirmed that no excess water was present outside the pores at or below liquid loading of 0.60 cm$^3$/g silica, and that the cubic ice I*c* is therefore formed entirely within the silica pores. The measured freezing and melting points of the confined cubic ice I*c* are 225 K and 230 K, respectively, consistent with the findings of Tonauer *et al*[36] (Extended Data Fig. 3). At larger liquid loadings of 0.70 cm$^3$/g silica and 0.94 cm$^3$/g silica, additional sharp Bragg peaks characteristic of ice I*h* emerge, consistent with previous studies[32,33]. At these higher loadings, DSC shows a feature corresponding to the formation of bulk $D_2O$ ice I*h* with a melting point of 276.9 K[37]. At all loading ratios, there is no signal from the THF structure II clathrate hydrate or, indeed, crystalline THF.

To determine the detailed crystal characteristics of the observed ice polymorph, the diffraction patterns for samples at liquid loadings of 0.34 cm$^3$/g, 0.60 cm$^3$/g, and 0.70 cm$^3$/g silica were then analyzed with the Rietveld method[38] using the GSAS-II[39] software package (liquid loading 0.60 cm$^3$/g silica, Fig. 2; liquid loading 0.34 cm$^3$/g and 0.70 cm$^3$/g silica, Extended Data Fig.

4). For a liquid loading of 0.60 $cm^3$/g silica the Bragg peaks corresponding to the ice crystallites are index-matched solely with pure cubic ice I*c*, symmetry $Fd\bar{3}m$ (Extended Table 1), with a weighted profile agreement-index $R_w$ of 4.39% and $R_{bkg}$ of 4.27%. The overall Bragg reflection peak-shape is consistent with, for example, a signature single component peak at 1.70 $Å^{-1}$ arising from the cubic ice I*c* (111) reflection. The peak widths are consistent with ice formation within the pores, yielding equatorial and axial dimensions of approximately 3 nm and > 6 nm, respectively. The latter axial crystallite dimension increases with pore loading while maintaining the cubic ice I*c* structure. We further verified the high structural purity of the cubic ice I*c* by attempting to refine the data against (i) pure ice I*h* and (ii) ice I*c*/I*h* mixture models, confirming no detectable stacking disorder within the instrument resolution, and by independently checking the diffuse scattering background (Extended Data Fig. 5; refined parameters in Table 1). From the fitted parameters, it is clear that the cubic ice I*c* crystallites grow primarily along the axial direction as the loading increases from 0.34 to 0.60 $cm^3$/g silica loadings; at 0.70 $cm^3$/g silica loading, hexagonal ice I*h* begins to form outside the pores.

Beyond the cubic ice I*c* Bragg pattern, the broad liquid-like scattering features indicate the presence of a non-crystalline component in the system at 150 K (Fig. 2). Previous studies[34] suggest that this corresponds to confined residual liquid that will eventually freeze at sufficiently low temperature. Combining these diffraction features with Raman spectroscopy, we discern that the unfrozen component near the MCM-41 pore surface is most likely to be a glassy THF-water phase (Extended Data Fig. 6), with THF molecules residing in this interfacial region. In the presence of THF molecules, the donor-acceptor (DA) water Raman peak red-shifts (~11.6 $cm^{-1}$) and broadens (FWHM ~13.2 $cm^{-1}$, +5%), alongside an increased DA water fraction and an intensity reduction at 980 $cm^{-1}$ (Si-O stretching vibrations of Si-OH groups), compared to neat $D_2O$ in MCM-41 (Extended Data Fig. 7). These spectroscopic signatures collectively confirm the interactions between THF and surface silanol groups.

In addition, quasi-elastic neutron scattering (QENS) elastic fixed-window scans (Methods, Extended Data Fig. 8) reveal that at the molecular scale the THF dynamics are coupled to those of the water, since they gradually arrest on cooling, but at temperatures higher than the bulk THF freezing point (167.4 K). Thus, we conclude that the non-crystalline THF-water component forms an amorphous glassy phase that is adjacent to crystalline cubic ice I*c* within the same pore. Prior to the formation of these ice phases, the precursor liquid phase was cooled from ambient temperature and the water dynamics extracted from QENS spectra (Methods). The quasi-elastic signal was deconvolved into two distinct motions which were observed in the temporal window of the spectrometer at different temperature ranges. These motions were isotropic rotation of H-atoms around the centre-of-mass of water molecules (240 - 260 K) and H jumps which describe the translational diffusion of water molecules (240 - 300 K). The former is characterised by its rotational residence time ($\tau_{rot}$) and the latter is described by its translational residence time ($\tau_{trans}$), jump length ($L$) and diffusion coefficient ($D_{trans}$). Notably, all these dynamical parameters are distinct from bulk water (Extended Data Fig. 9d) and those of water confined in mesoporous silicas of various pore sizes. Thus, the inclusion of a clathrate-forming molecule imposed unique water dynamics in the liquid phase upon its approach to pristine cubic ice I*c*.

To assess the specific role of clathrate-forming THF molecules in the formation of pure confined ice I*c*, we collected neutron diffraction patterns for two control samples using Polaris: (i) neat $D_2O$, and (ii) acetonitrile-$D_3$/$D_2O$ mixture at 1:17 molar ratio, both confined in MCM-41 matrix at an identical liquid loading of 0.60 $cm^3/g$ silica (Fig. 3a). Here, acetonitrile serves as a non-hydrate forming chemical analogue of THF, being a water soluble polar aprotic molecule, known to segregate near hydroxylated silica surfaces at high water fraction[40]. Clearly, there are degrees of scattering from hexagonal/stacking disorder in both controls, with, for example, the (111) ice I*c* peak at 1.70 $Å^{-1}$ becoming split into a triplet of distinct features at 1.61 Å$^{-1}$, 1.71 Å$^{-1}$, and 1.82 Å$^{-1}$ corresponding to the signature (100), (002), and (101) reflections of ice I*h*. Differential scanning calorimetry (DSC) thermograms show that in all cases the ice phases are only forming within the MCM-41 nanopores (Fig. 3b). The observation of hexagonal ice in the presence of acetonitrile confirms that the presence of clathrate-forming THF molecules plays a causal role in the formation of pristine cubic ice I*c*, rather than a confinement effect alone.

To corroborate the cubic ice I*c* crystallinity and hydrogen-bond network along with the interfacial location of THF molecules, we performed Raman spectroscopy analysis of the 1:17 molar ratio THF-water mixture at liquid loading 0.60 $cm^3/g$ silica (Fig. 4). We used $D_2O$ and THF-$H_8$/$D_2O$ here, with isotopic substitution in the THF expected to have minimal impact on the ice structure[41]. Lorentzian fitting of the water spectra shows that the O-D stretching band (Fig. 4a) can be deconvolved into four distinct components: three corresponding to the main O-D stretching vibrations[42] within the THF-$H_8$/$D_2O$ mixture, and one from the MCM-41 matrix. The low-frequency component at 2325 $cm^{-1}$ is associated with the formation of a tetrahedral hydrogen-bonding network (termed DDAA where, as before, D is a hydrogen-bond donor and A is a hydrogen-bond acceptor), and is assigned as an ice-peak[43,44]. The higher-frequency components are associated with 2-coordinated hydrogen-bonded water (DA) and the O-D stretch free from hydrogen-bonding (free-OD), respectively. For simplicity, negligible effects of 3-coordinate DAA and DDA are omitted. The DDAA component narrows (FWHM ~ 4.3 $cm^{-1}$, a -6% change) on going from stacking disordered ice ($D_2O$ in MCM-41) to cubic ice I*c* formed in the presence of THF molecules (Fig. 4b). The narrowing indicates the loss of stacking disorder and formation of more ordered single-phase ice crystallites in the THF-$H_8$/$D_2O$ at liquid loading 0.60 $cm^3/g$ silica, consistent with previous Raman spectroscopy studies in the bulk state[25,45].

In addition to fundamental interest in the formation and stability of ice phases, there are numerous direct implications for the discovery of a liquid-based route to thermodynamically stable cubic ice I*c*. For example, cirrus and polar-mesospheric clouds (PMCs) nucleate on mineral-organic aerosols[46] at temperatures consistent with our observations, where organic aerosols frequently vitrify and act as efficient ice-nucleating particles[12]. This is precisely the confinement-organic synergy that our strategy exploits. For PMC systems, our findings support the hypothesis that the primary crystalline ice phase can be cubic[4,5], challenging the recent view that upper-atmosphere ice is predominantly stacking-disordered[47]. While our choice of organic clathrate promoter may at first seem rather esoteric, recent *in situ* mass spectrometry of comet 67P reveals abundant oxygen-bearing heterocycles and identifies THF as a particularly promising target for interstellar searches[48]. The observation of cubic ice I*c* formation adjacent to an interfacial amorphous THF-water phase may help to explain why icy dust grains in comets and the interstellar medium can preserve mixed cubic-amorphous mantles[5]. Moreover, we find that nanoconfinement suppresses clathrate formation relative to the bulk, informing clathrate management in terrestrial contexts[49] and providing microphysical constraints on cometary-nucleus outgassing models that hinge on amorphous-crystalline-clathrate transitions[50].

In summary, we have shown that a simple combination of nanoconfinement and clathrate-promoting organic molecules delivers stable, reproducible cubic ice I*c* without stacking disorder directly from the liquid state at ambient pressure and temperatures as high as 230 K. Control experiments show that the combination of nanoconfinement and the presence of a clathrate (water-cage) promoter is key. This approach does not require extreme cryogenic conditions, high-pressure precursors, or complex gas-hydrate intermediates, and resolves the longstanding quest for a liquid-based route to pristine cubic ice 1*c*. The discovery that cubic ice I*c* is directly accessible from the liquid state strongly suggests that this polymorph plays a much more significant role in natural and synthetic processes than previously thought, across a broad range of terrestrial and extraterrestrial environments.

**Fig. 1: Neutron diffraction patterns showing cubic ice I*c* formation in mesoporous silica at ambient pressure.**

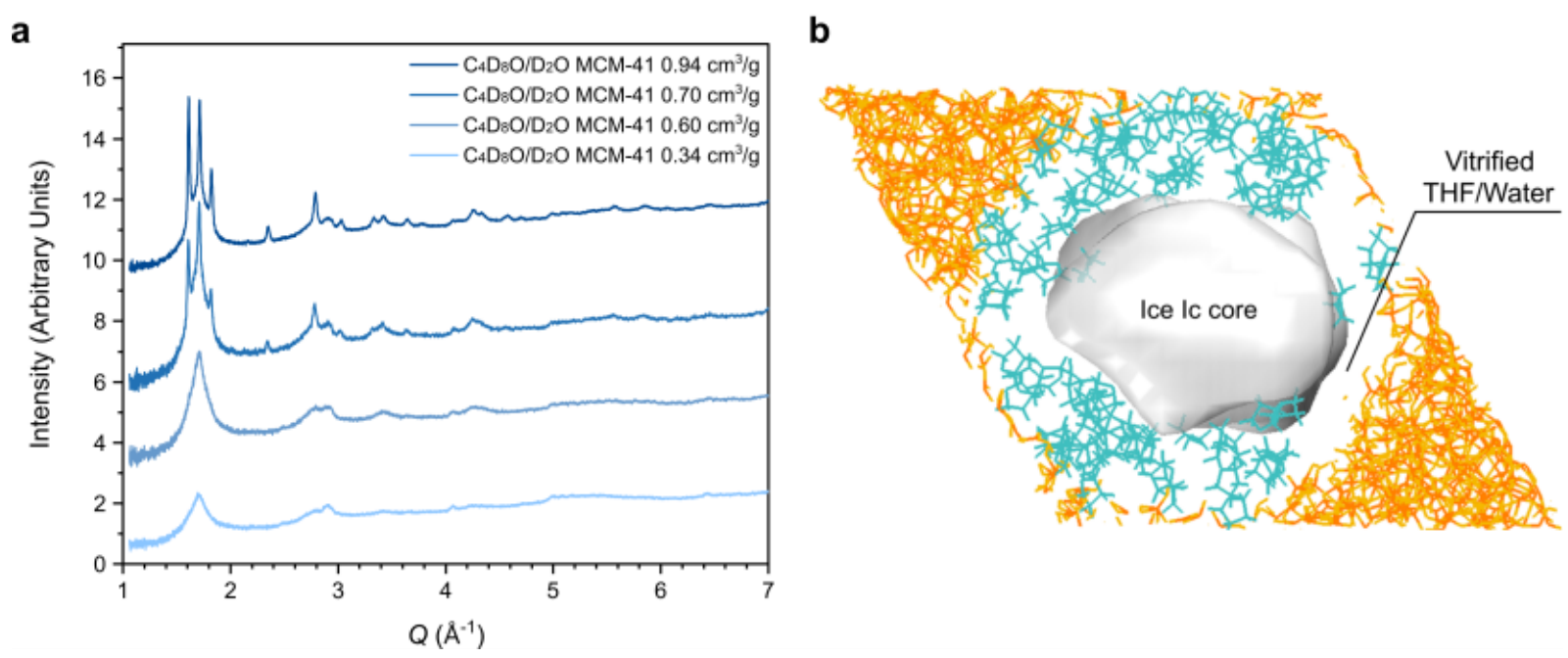

**a**. Neutron diffraction patterns collected by the Polaris neutron diffractometer (ISIS Neutron and Muon Source) from samples of deuterated 1:17 molar ratio mixture of tetrahydrofuran-water ($C_4D_8O/D_2O$) confined in MCM-41 silica at ambient pressure. Data collected at 150 K, showing formation of pristine cubic ice I*c* at liquid loadings of 0.60 and 0.34 $cm^3$/g silica, as seen through clear Bragg peaks emerging at momentum transfer $Q$ = 1.70 $Å^{-1}$, 2.78 $Å^{-1}$, 3.42 $Å^{-1}$, 4.28 $Å^{-1}$; the unique fingerprint of cubic ice I*c*[25,26]. At liquid loadings of 0.70 and 0.94 $cm^3$/g silica, the triplet peak at 1.5 - 2.0 $Å^{-1}$ indicates the appearance of ice crystallites other than cubic ice I*c*, for example hexagonal I*h* or stacking disordered I*sd*. Additional Bragg peaks are observed in the sample with liquid loading 0.70 $cm^3$/g silica at 1.61 $Å^{-1}$, 1.71 $Å^{-1}$, 1.82 $Å^{-1}$, 2.34 $Å^{-1}$, 2.78 $Å^{-1}$, 3.03 $Å^{-1}$, 3.32 $Å^{-1}$ and in the sample with liquid loading 0.94 $cm^3$/g silica at 1.61 $Å^{-1}$, 1.71 $Å^{-1}$, 1.82 $Å^{-1}$, 2.35 $Å^{-1}$, 2.79 $Å^{-1}$, 3.03 $Å^{-1}$, 3.33 $Å^{-1}$. These peaks correspond to the (100), (002), (101), (102), (110), (103), (112) reflections of hexagonal ice I*h*[31]. No reflections attributed to THF-water hydrate formation are seen across all filling ratios. All patterns are vertically offset for clarity. **b**. Schematic illustration of the cooled confined tetrahydrofuran-water, with a core of ice I*c* surrounded by a tetrahydrofuran-rich glass at the silica interface. The box-side is 4.33 nm and pore radius 1.87 nm.

**Fig. 2: Rietveld refinement to the neutron diffraction pattern of cubic ice I*c* formed in mesoporous silica at a liquid loading of 0.60 cm$^3$/g silica.**

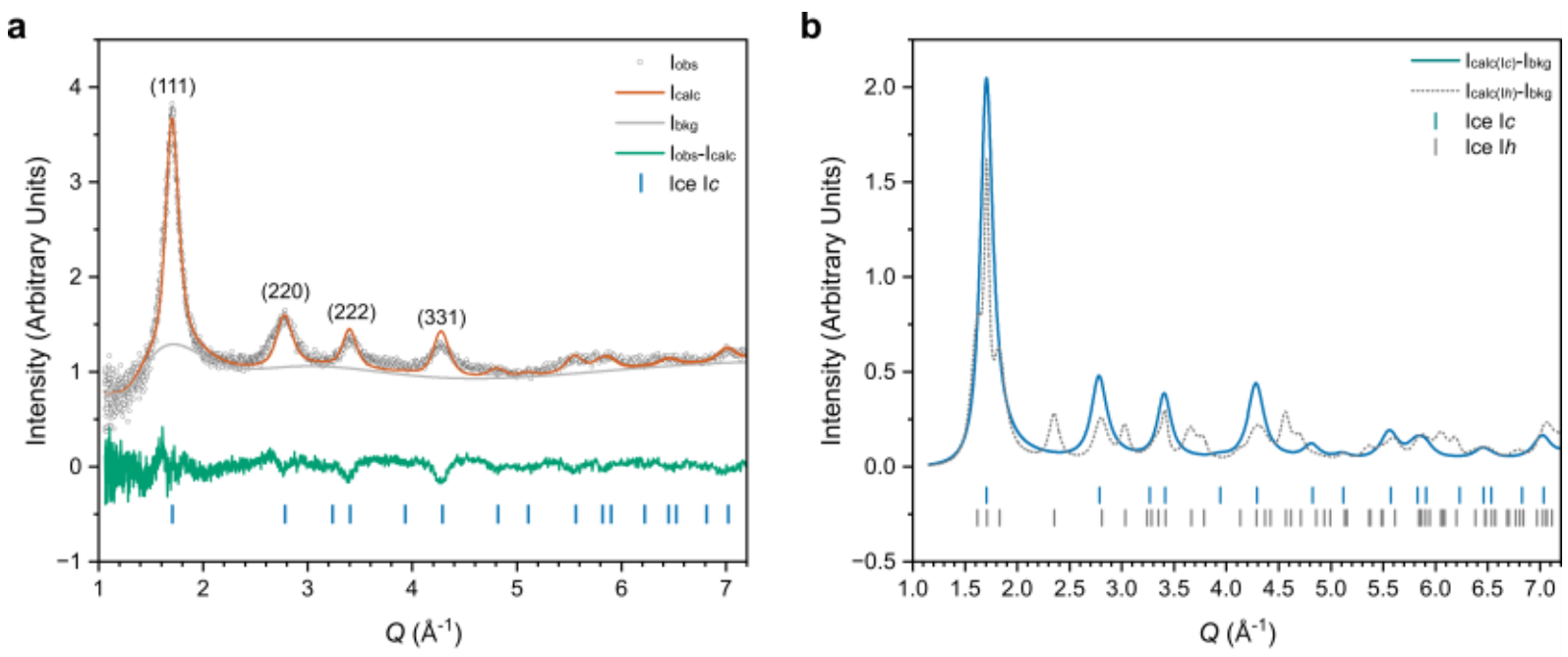

**a.** Rietveld refinement to pristine cubic ice I*c* (red line) agrees closely with the experimental data $I_{obs}$ (circled points) collected for a deuterated 1:17 molar ratio mixture of tetrahydrofuran-water ($C_4D_8O/D_2O$) confined in MCM-41 silica at a liquid loading of 0.60 cm$^3$/g silica. Experimental data were collected at ambient pressure and 150 K. The residuals $I_{obs}$–$I_{calc}$ (green line) and the positions of the ice I*c* peaks (blue bars) are shown in the panel. The background $I_{bkg}$ (grey line) accounts for the diffuse scattering from the amorphous fraction within the pores. Fit parameters are given in Table 1. **b**. Comparison of Rietveld refined profiles for pure cubic ice I*c* (blue line, peak positions blue bars) and pure hexagonal ice I*h* (grey line, peak positions grey bars), showing that the experimental data can only be fitted to cubic ice I*c*.

**Fig. 3: Control experiments showing that pristine cubic ice I*c* only crystallizes in the presence of a clathrate-hydrate promoting molecule (tetrahydrofuran, THF)**

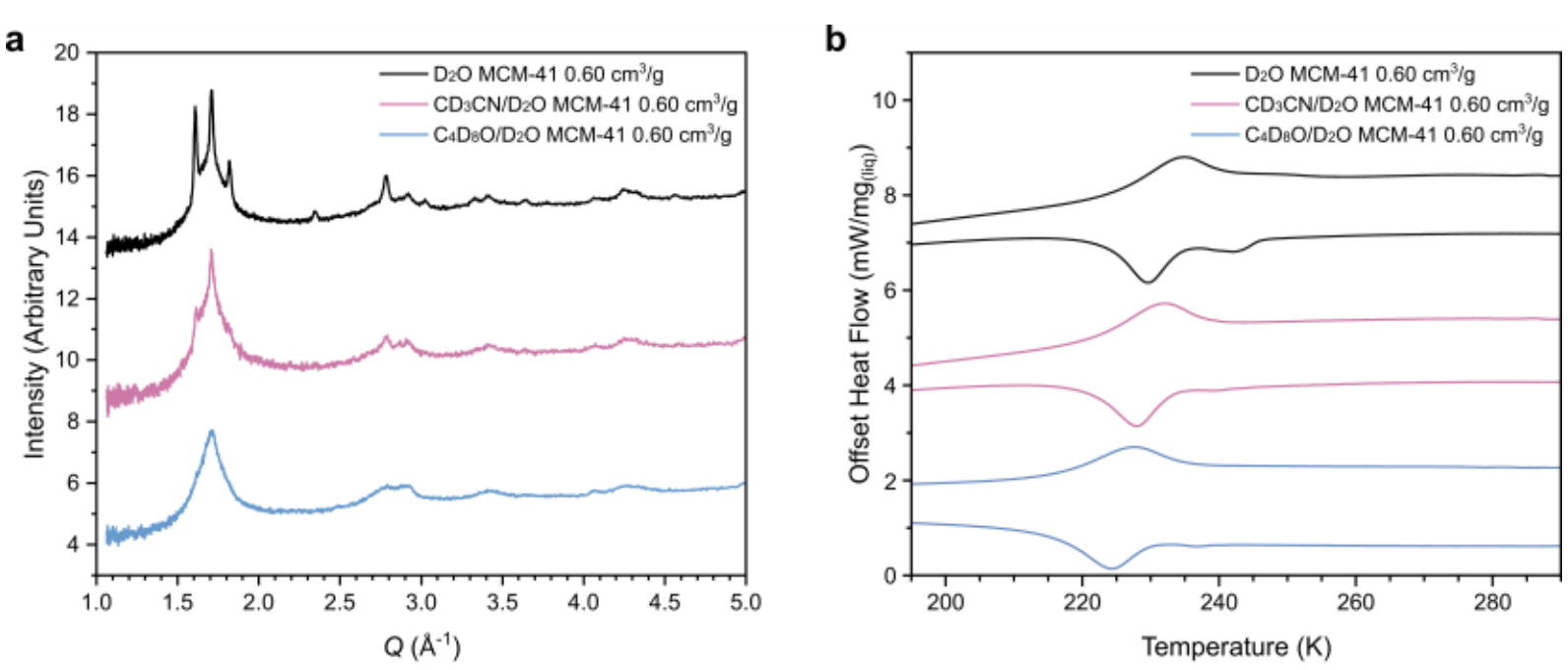

**a.** Neutron powder diffraction patterns collected by the Polaris diffractometer (ISIS Neutron and Muon Source) at 150K. The cubic ice I*c* forming 1:17 molar ratio mixture of THF-water ($C_4D_8O/D_2O$) in MCM-41 (blue line). Control experiments on neat $D_2O$ in MCM-41 (black line), and 1:17 molar ratio mixture of acetonitrile-water ($CD_3CN/D_2O$) in MCM-41 (pink line). All samples studied at a liquid loading of 0.6 cm$^3$/g silica. Note that both neat $D_2O$ and $CD_3CN/D_2O$ samples show clear triplet peak patterns at $Q$ = 1.61 Å$^{-1}$, 1.71 Å$^{-1}$, 1.82 Å$^{-1}$, and also at 2.78 Å$^{-1}$, diagnostic of ice I*h*/I*sd* rather than a single-phase cubic ice I*c* obtained from the $C_4D_8O$-$D_2O$ mixture. All data are vertically offset for clarity. **b.** Corresponding differential scanning calorimetry (DSC) thermograms showing that the ice phases are only forming within the MCM-41 nanopores. Further DSC thermograms are provided in Extended Data Fig. 3. We further note that a small exothermal peak appears at around 245 K in the $D_2O$ MCM-41 sample, which we propose arises from water located near the pore mouths.

**Fig. 4: Raman characterization of the OD stretch in cubic ice I*c*.**

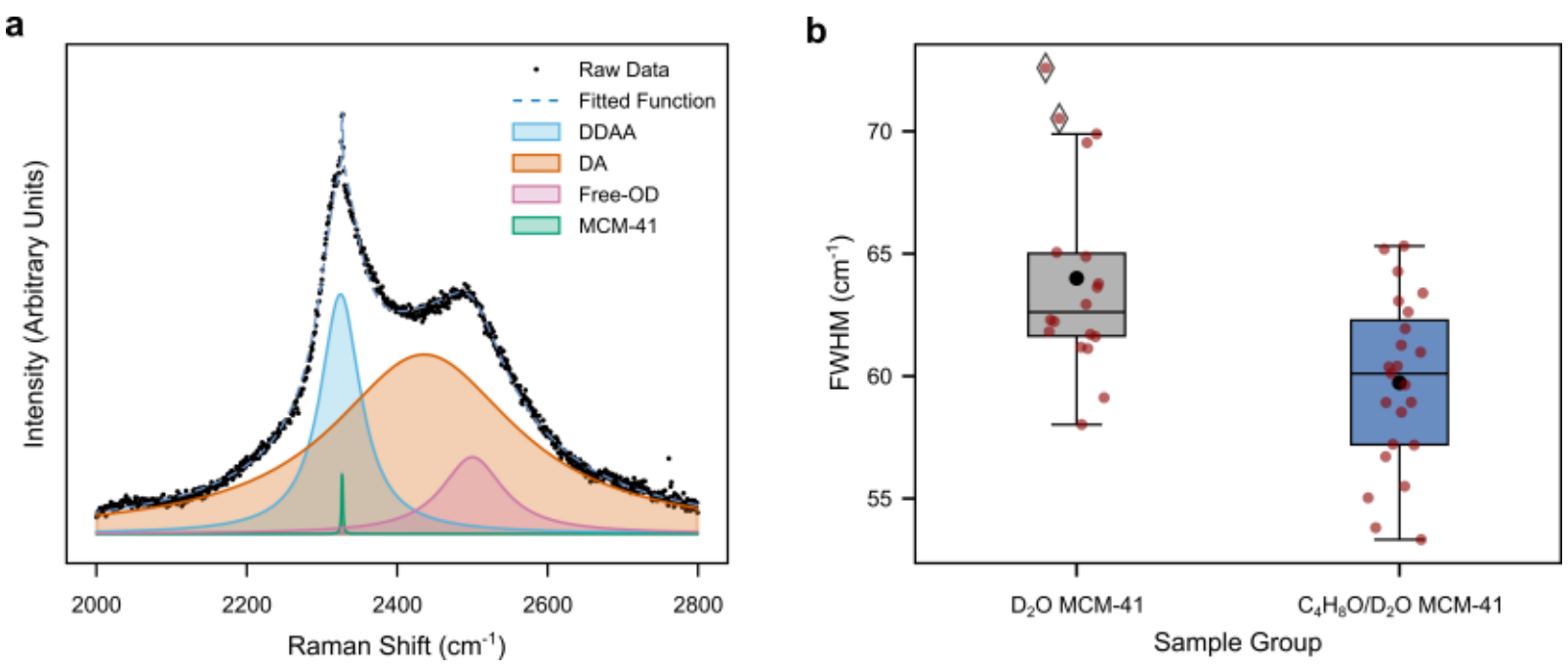

**a**. Representative Raman spectrum (black line) at 150 K for cubic ice I*c* formed from nanoconfined $C_4H_8O/D_2O$ mixture (1:17 molar ratio) at a liquid loading of 0.60 $cm^3$/g silica. Lorentzian fits of the OD-stretching bands (around 2325 $cm^{-1}$) for the hydrogen bonding (HB) configuration are shown in blue, orange and pink, respectively; DDAA (D: HB donor, A: HB acceptor), DA and free-OD. The small green peak is attributed to MCM-41. **b.** Full-width at half-maximum (FWHM) values extracted from the DDAA component of Raman spectra, comparing the OD stretch in neat $D_2O$ at (grey) and 1:17 molar ratio $C_4H_8O$-$D_2O$ mixture (blue) at 0.60 $cm^3$/g silica loading. Red dots indicate individual spectral measurements (n = 14 points per group); Horizontal lines in the box indicate the median values, and black points represent mean values. Grey diamonds represent the outliers of the data, which were not considered for mean and median value calculation. Confined cubic ice I*c*, formed from the $C_4H_8O$-$D_2O$ mixture, therefore exhibits a subtle narrowing of the OD-stretch with the FWHM reduced by ~4.3 $cm^{-1}$ (mean value), indicating a greater lattice order relative to the stacking disordered ice I*sd* formed from confined neat $D_2O$.

## Methods

### Sample preparation

Adsorption isotherms using the Brunauer-Emmett-Teller method on MCM-41 yield pore volume and pore radius of 0.78 $cm^3/g$ and 1.87 nm respectively. Deuterated THF ($C_4D_8O$) was purchased from Fluorochem Ltd (≥ 99.5 atom % D). Deuterated acetonitrile ($CD_3CN$) was purchased from Sigma Aldrich (≥ 99.8 atom % D). Deuterated water was from Thermo Fisher Scientific. Protiated THF ($C_4H_8O$) was purchased from Sigma Aldrich. Milli Q pure water was used as protiated water. All liquids were ≥ 99.5% pure. MCM-41 samples were outgassed in Schlenk line (room temperature, $<10^{-1}$ mbar) for three days prior to use. After adding liquids to the silica, samples were sealed in air-tight vials and left for at least 3 hours to equilibrate before any measurements.

### Neutron diffraction

Time-of-flight powder diffraction data were acquired on Polaris at the ISIS Neutron and Muon Source using five detector banks. The banks cover $2\theta$ ranges of 6.7-14.0° (bank 1), 19.5-34.1° (2), 40.4-66.4° (3), 75.2-112.9° (4) and 134.6-167.4° (5), with $\Delta d/d \approx 2.7$ %, 1.5%, 0.85%, 0.51%, 0.30% respectively[51]. Samples were loaded into 8mm diameter thin-walled cylindrical vanadium containers and analyzed with 40 mm height × 15 mm width beam. Samples were cooled down to 150 K using a standard orange cryostat at a cooling rate approximately 10 K/min. Different pore loadings of $C_4D_8O/D_2O$ samples were measured on the neutron beam for a minimum of 3 h each. Control measurements were performed on neat $D_2O$ and $CD_3CN/D_2O$ (1:17 molar ratio) samples, both at a liquid loading of 0.60 $cm^3/g$ silica. The samples were exposed to the neutron beam for 2.5 h and 3 h, respectively. Deuterated liquids were employed to reduce incoherent scattering contributions from protium. In addition to the samples themselves, scattering data were collected on the neutron beam at 150 K for a minimum of 3 h each for: the empty instrument fitted with the cryostat, empty 8 mm vanadium containers in the cryostat, and evacuated MCM-41 loaded in 8 mm vanadium containers in the cryostat.

Raw time-resolved data were imported into Mantid[52], corrected for detector efficiency, time-focused into *d*-spacing and *Q*-spacing, background-subtracted, and then exported as flux-normalized diffraction patterns ready for Rietveld analysis. The Polaris instrumental calibration file was obtained by measuring silicon standard powder (NIST 640e Si) in the same cycle as the experiment. Rietveld refinement[38] was conducted using GSAS-II software package[39], using datasets collected on the bank 3. TOF profile function 3 is used to reproduce the characteristic diffraction peaks. It convolves a pseudo-Voigt line shape with two opposing exponential decay function.

The cubic ice I*c* reproducibility experiment was done on the OSIRIS spectrometer at ISIS by using its high-resolution diffractometers. Data was collected using *drange* 1-5 (*Q*-space 8.9771-1.0135 $Å^{-1}$, $5 \times 10^{-3} < \Delta d/d < 6 \times 10^{-3}$). $C_4D_8O/H_2O$ and $C_4H_8O/D_2O$ at liquid loading 0.60 $cm^3/g$ silica were prepared in the same way as discussed in 'sample preparation' and transferred into their respective sachets of aluminium foil (5 cm × 7 cm × 1 mm). The sachet was sealed and the sample was evenly distributed within it. Each sachet was subsequently inserted around the central annulus of an aluminum can (height 50 mm), where it occupied the 2 mm gap between the annulus and the outer walls of the can. Each sample can was subsequently sealed with indium wire, analyzed with 44 mm height × 22 mm width beam. Samples were cooled down from 300 K to 5 K using a cryostat available at OSIRIS, at a cooling rate of approximately 1.5 K/min. Raw data were collected in five separate but slightly overlapped TOF channels (available in ISIS data repository). Data processing was also performed in Mantid, including time-focusing into *Q*-spacing, vanadium normalization, and merging the five TOF windows into a continuous pattern. Unlike Polaris data, any slight mismatches in normalization or background subtraction between adjacent *dranges* on OSIRIS would produce artifacts at the bin boundaries.

## Neutron spectroscopy

Quasi-elastic neutron scattering (QENS) experiment was performed on the OSIRIS beamline, ISIS Neutron and Muon Source. The PG002 configuration was used to examine energy transfers over a $Q$-range 0.24 to 1.80 Å$^{-1}$, with energy resolution ($\Delta E$) of 25 µeV and a dynamic range of -0.5/+1.0 meV. All QENS data was analyzed in Mantid[52]. Elastic fixed window scan (EFWS) was performed on both $C_4D_8O/H_2O$ and $C_4H_8O/D_2O$ at liquid loading 0.60 cm$^3$/g silica from 300 K-150 K to locate transitions for both water and THF, respectively. EFWS data was produced by integrating neutron counts between $\Delta E$ ranges of ± 0.0125 meV and summed across the whole $Q$-range of 0.24 to 1.80 Å$^{-1}$. For QENS analysis, the $C_4D_8O/H_2O$ at liquid loading 0.60 cm$^3$/g silica sample was used to reveal the dynamics of the supercooled water during its phase transition to cubic ice I*c*, since incoherent scattering is dominated by the protium. QENS measurements were taken at 5 K, 240 K, 260 K, 280 K and 300 K, with 4.5 hours counting time for each temperature point. Detectors were grouped into 14 $Q$ groups. Analysis utilized the $Q$-range 0.24 to 1.80 Å$^{-1}$ and a $\Delta E$ range -0.3 to +0.6 meV, with constant bin widths of 0.0005 meV. The reduced QENS data comprises an incoherent component, arising from self-correlations of individual nuclei, and a coherent component, reflecting spatial correlations between different nuclei; both convolved with the instrument's energy resolution. In the modelling, any motions slower than the experimental time window and coherent structural components would be represented as a delta function, while any motions faster than the experimental time window would be represented as a flat background. The QENS spectrum at the base temperature (5 K) is a convolution of the delta function (i.e., elastic scattering contributions and inelastic contributions slower than the experimental window) and instrumental resolution. The remaining incoherent contributions can be decoupled into translational and rotational components, treating each as independent processes and representing them with Lorentzian functions.

The translational broadening in QENS, arising from center-of-mass motions, has a linewidth $\Gamma_{\mathrm{trans}}$ (half-width at half-maximum, HWHM) that scales with $Q^2$. By fitting this dispersion with a jump-diffusion model, where molecules vibrate within a site for an average residence time $\tau_{\mathrm{trans}}$ before making a rapid jump of length $L$ over a much shorter time $\tau_1$ (with $\tau_{\mathrm{trans}} \gg \tau_1$), we can determine the translational diffusion coefficient, $D_{\mathrm{trans}}$. The best fit was to the Teixeira water model[53];

$$\Gamma_{trans}(Q) = \frac{\hbar}{\tau_{trans}} \cdot \frac{(QL)^2}{6+(QL)^2}, \tag{1}$$

$$D_{trans} = \frac{L^2}{6\tau_{trans}}. \tag{2}$$

While the translational broadening is in the experimental time window across all measured QENS temperature points, rotational broadening is only captured at and below 260 K. The rotational term is modelled first as a Lorentzian with a $Q$-independent linewidth $\Gamma_{\mathrm{rot}}$ (HWHM). The rotational model is then refined by fixing the translational terms and applying an isotropic rotational-diffusion model that describes the H atoms reorienting about the molecular centre of mass. For continuous isotropic rotations of water molecules, the rotational dynamical structure factor is given by the Sears expansion[54];

$$S_{rotation}(Q,\omega) = j_0^2(aQ)\delta(\omega) + \sum_{l=1}^{\infty}(2l+1)\frac{j_l^2(aQ)}{\pi}\frac{l(l+1)\left(\frac{1}{\tau_{rot}}\right)}{\left[l(l+1)\left(\frac{1}{\tau_{rot}}\right)\right]^2+\omega^2}. \quad (3)$$

Here, $j_l(x)$ are the spherical Bessel functions and the radius of rotation $a$ was fixed at 0.98 Å (OH bond length in water molecules), $\tau_{\mathrm{rot}}$ is the rotational relaxation time.

## Raman measurement

Raman maps (to collect multiple spectra from a particular region) and single point spectra were acquired using a Renishaw in Via confocal micro-Raman instrument fitted with a 514.5 nm diode laser calibrated by a Si wafer. The temperature control was achieved by a liquid nitrogen–based Linkam THMS 600 cooling stage, operating between 120 and 300 K. The prepared samples were enclosed in a glass sample cell and mounted in the stage, surrounded by a nitrogen atmosphere. One cooling cycle was performed at each point at a cooling rate of 5 K/min. The spectra were then fitted using Lorentzian functions.

## DSC

Differential Scanning Calorimetry (DSC) was performed on a PerkinElmer DSC 8000 equipped with a liquid-nitrogen cooling system, over the range 190 - 300 K, with an empty sealed aluminium pan as reference. Helium was used as the purge gas at 20 mL/min. Each cycle began with an isothermal hold at 300 K for 1 min, followed by cooling to 190 K at 10 K/min and reheating to 300 K at 10 K/min.

## Acknowledgements

We thank the UK Science and Technology Facilities Council (STFC) for the award of beamtime and access to instruments at the ISIS Neutron and Muon Source (awards RB2310479 and RB2420268), including the Hydrogen and Catalysis Laboratory, facilitated by Dr James Taylor, and the Materials Characterisation Laboratory, facilitated by Dr Daniel Nye and Dr Gavin Stenning. M.N and A. J. C. thank the Royal Society for support via the University Research Fellowship scheme (URF\R1\221476). M.L.B. acknowledges support of the Engineering and Physical Sciences Research Council (EPSRC) through a PhD studentship (EP/W524335/1). C.D.M. thanks the Glasstone Bequest for support through the Violette and Samuel Glasstone Fellowships in Science Scheme.

## Author Contributions

N.T.S. and C.Z. conceptualized the study and designed the experiment. C.Z., P.F.H., M.N., C.D.M., A.J.C. and N.T.S. acquired the data in the Polaris experiment. C.Z., A.I.S., S.M., M.N. and M.F. acquired the data in the OSIRIS experiment. Y.W. and C.Z. carried out the Raman experiments. M.L.B., C.G.S. and A.J.C carried out the DSC experiments. C.Z. performed the diffraction data analysis. C.Z. and A.I.S. performed the QENS data analysis. Y.W., C.Z. and C.A.H. performed the Raman data analysis. M.L.B. and A.J.C. performed DSC data analysis. T.F.H. and M.F. contributed to preliminary neutron experiments. P.F.H., S.M., C.A.H. and C.G.S validated the data analysis. C.Z., Y.W. and A.J.C. prepared the figures and visualization. C.Z., A.I.S. and N.T.S. wrote the original draft. All the authors reviewed and edited the paper.

## Competing interests

The authors declare no competing interests.

**Table 1. Structural parameters of ice I*c* and I*h* as obtained by the Rietveld refinement to neutron diffraction data.**

| Sample | Ice phase (*Space group*) | Phase fraction | *Rw* (%) | *R* (%) | $R_{bkg}$ (%) | $RF^2$ (%) | GOF | Equatorial size (nm) | Axial size (nm) | Unit cell (Å) | Atom | *U*iso (Å$^2$) | Occ. |
|---|---|---|---|---|---|---|---|---|---|---|---|---|---|
| $C_4D_8O/D_2O$ MCM-41 0.34 $cm^3/g$ | I*c* *Fd*-3*m* | 100% | 9.497 | 8.96 | 12.14 | 17.25 | 1.87 | 3.47 | 6.7 | ***a*** 6.37967 | O | 0.02373 | 1.0 |
| | | | | | | | | | | | D1 | 0.03313 | 0.5 |
| | | | | | | | | | | | D2 | 0.03313 | 0.5 |
| $C_4D_8O/D_2O$ MCM-41 0.60 $cm^3/g$ | I*c* *Fd*-3*m* | 100% | 4.388 | 4.22 | 4.27 | 18.37 | 1.68 | 3.54 | 12.0 | ***a*** 6.38764 | O | 0.02384 | 1.0 |
| | | | | | | | | | | | D1 | 0.03004 | 0.5 |
| | | | | | | | | | | | D2 | 0.03004 | 0.5 |
| $C_4D_8O/D_2O$ MCM-41 0.70 $cm^3/g$ | I*c* *Fd*-3*m* | 71.505% | 5.045 | 4.87 | 4.91 | 10.77 | 1.77 | 3.54 | 12.0 | ***a*** 6.39506 | O | 0.03132 | 1.0 |
| | | | | | | | | | | | D1 | 0.03653 | 0.5 |
| | | | | | | | | | | | D2 | 0.03653 | 0.5 |
| | I*h* *P*6$_3$/*mmc* | 28.495% | | | | 17.81 | | 76.2 (isotropic model) | | ***a*** 4.51722 ***c*** 7.366 | O | 0.02000 | 1.0 |
| | | | | | | | | | | | D1 | 0.03347 | 0.5 |
| | | | | | | | | | | | D2 | 0.03347 | 0.5 |

## Extended data figures and table

**Extended Data Fig. 1: Total neutron diffraction patterns of ice formation in mesoporous silica MCM-41.**

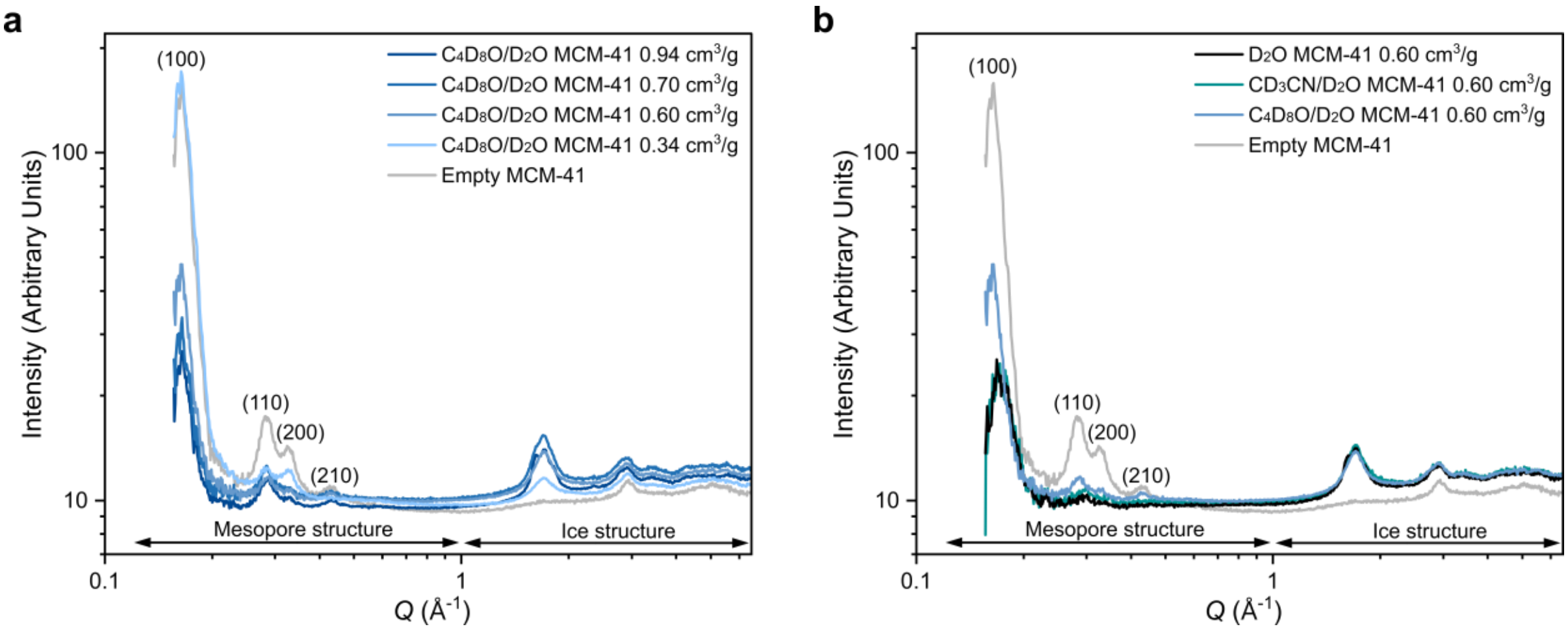

**a**. Total neutron diffraction patterns collected at Polaris diffractometer (ISIS). Samples are empty MCM-41 (grey line), and 1:17 molar ratio $C_4D_8O/D_2O$ within MCM-41 at liquid loadings of 0.34, 0.60, 0.70, 0.94 $cm^3/g$ silica (blue lines), measured over $0.16 \leq Q \leq 6.7$ Å$^{-1}$ at 150 K. **b.** Comparison of total neutron diffraction patterns of 1:17 molar ratio $C_4D_8O/D_2O$ with pure $D_2O$ (black line), 1:17 molar ratio $CD_3CN/D_2O$ (green line) all at liquid loadings of 0.60 $cm^3/g$ silica at 150 K. Mesopore structure: Low-*Q* (100), (110), (200), (210) Bragg peaks arise from the hexagonal lattice of the cylindrical pores, with intensities governed by the scattering contrast between the ice, amorphous component and pore walls. Ice structure: Intermediate and high-*Q* scattering is dominated by ice diffraction peaks. Note, to observe the MCM-41 Bragg peaks these measurements utilized the low-angle bank of Polaris which has a low-resolution (2.7%)[51].

**Extended Data Fig. 2: Neutron diffraction pattern from confined cubic ice I*c* using the OSIRIS beamline at 5 K.**

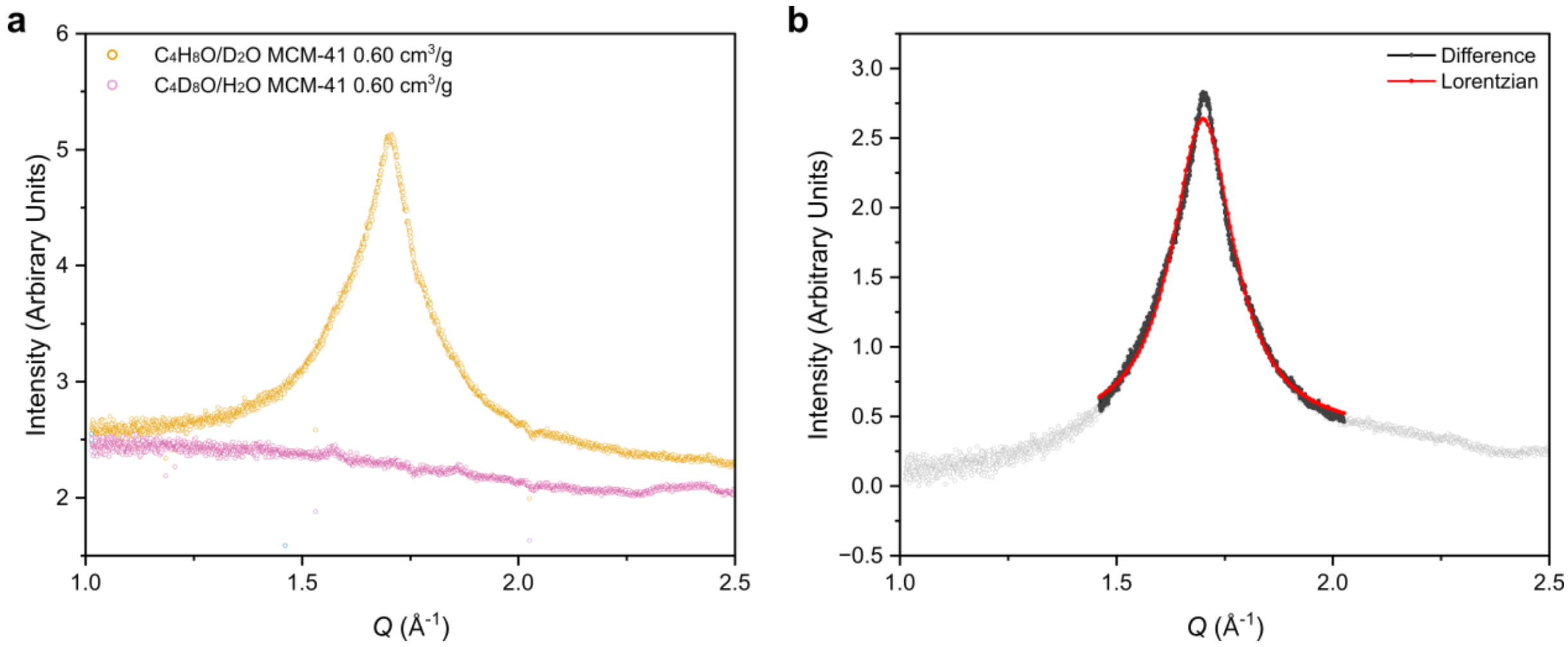

Contrast-variation neutron diffraction patterns were recorded on the high-resolution diffractometer setting of the OSIRIS beamline (ISIS Neutron and Muon Source, UK). The data are focused on the cubic ice I*c* (111) Bragg peak located at 1.70 $Å^{-1}$. **a.** 1:17 molar ratio $C_4H_8O/D_2O$ (orange) and $C_4D_8O/H_2O$ (pink) mixtures confined within porous silica MCM-41 at liquid loading 0.60 $cm^3/g$ silica at 5 K, 10 h after ice formation. The signal is dominated by the deuterated components, hence the orange trace reports primarily on the $D_2O$ cubic ice I*c* lattice, while the pink trace is dominated by the amorphous $C_4D_8O$-rich component. **b.** Difference between 1:17 molar ratio $C_4H_8O/D_2O$ and $C_4D_8O/H_2O$ at liquid loading of 0.60 $cm^3/g$ silica (black line), which can be fitted by a single Lorentzian (red line). Note that there is no signal from the triplet Bragg peaks in the *Q*-region associated with hexagonal ice I*h* or stacking disordered ice I*sd*.

**Extended Data Fig. 3: Differential Scanning Calorimetry (DSC) thermograms of confined $C_4D_8O/D_2O$ (1:17 molar ratio) aqueous solution during cooling/heating between 190 K to 300 K.**

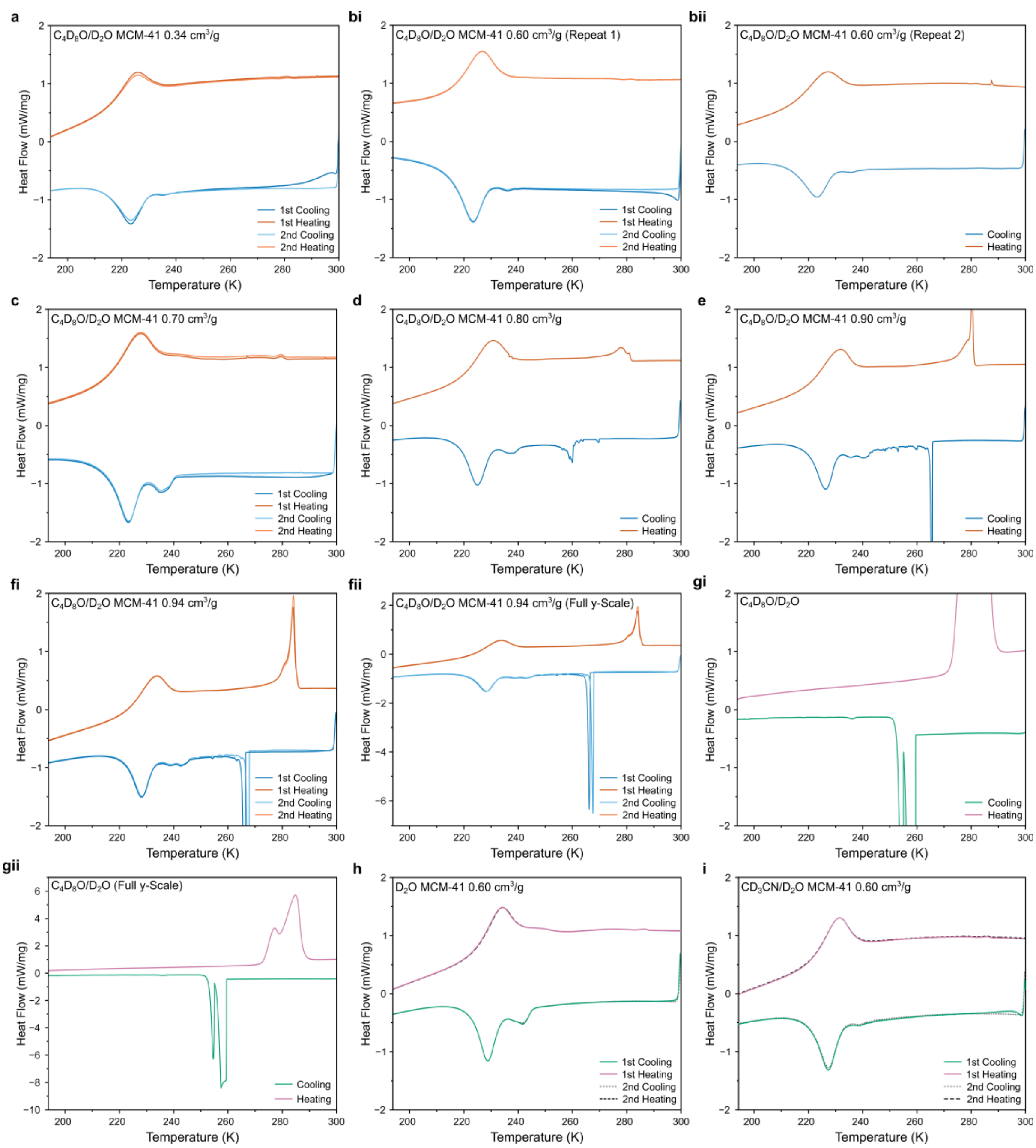

Differential Scanning Calorimetry measurements between 300 and 190 K were performed on (**a-f**) 1:17 molar ratio $C_4D_8O/D_2O$ at different liquid loadings in MCM-41 silica, (**g**) silica-free 1:17 molar ratio $C_4D_8O/D_2O$ (**h**) $D_2O$ at liquid loading 0.60 $cm^3/g$ silica in MCM-41, and (**i**) 1:17 molar ratio acetonitrile-water $CD_3CN/D_2O$ at a liquid loading 0.60 $cm^3/g$ silica in MCM-41. Data were normalized to the weight of liquid present. For the 1:17 molar ratio $C_4D_8O/D_2O$ at liquid loading ≥0.700 $cm^3/g$ silica, the DSC thermograms show a peak at 266 - 270 K during cooling, and a peak at 266 - 274 K during heating. These features are assigned to the bulk-like water outside the pores[37]. In contrast, no transition is observed near 270 K for samples with liquid loading ≤0.60 $cm^3/g$ silica, confirming the absence of bulk water. Instead, a broad peak centered at 225 K during cooling and at 230 K during heating is assigned to the ice formation inside the MCM-41 pores. No additional melting peaks are detected, indicating that the remaining tetrahydrofuran-water component does not crystallize. Melting points taken from onset, measured via intersection of linear fits of leading baseline and peak are (**a**) 215 K, (**bi**) 216 K, (**bii**) 215 K, (**c**) 214 K, (**d**) 219 K & 269 K, (**e**) 220 & 274 K, (**f**) 223 K & 277 K, (**g**) 273 K & 272 K, (**h**) 224 K, (**i**) 221 K.

**Extended Data Fig. 4: Rietveld refinement of neutron diffraction patterns recorded for Ice I*c* in MCM-41 at liquid loadings of 0.34 and 0.70 cm$^3$/g silica at 150 K.**

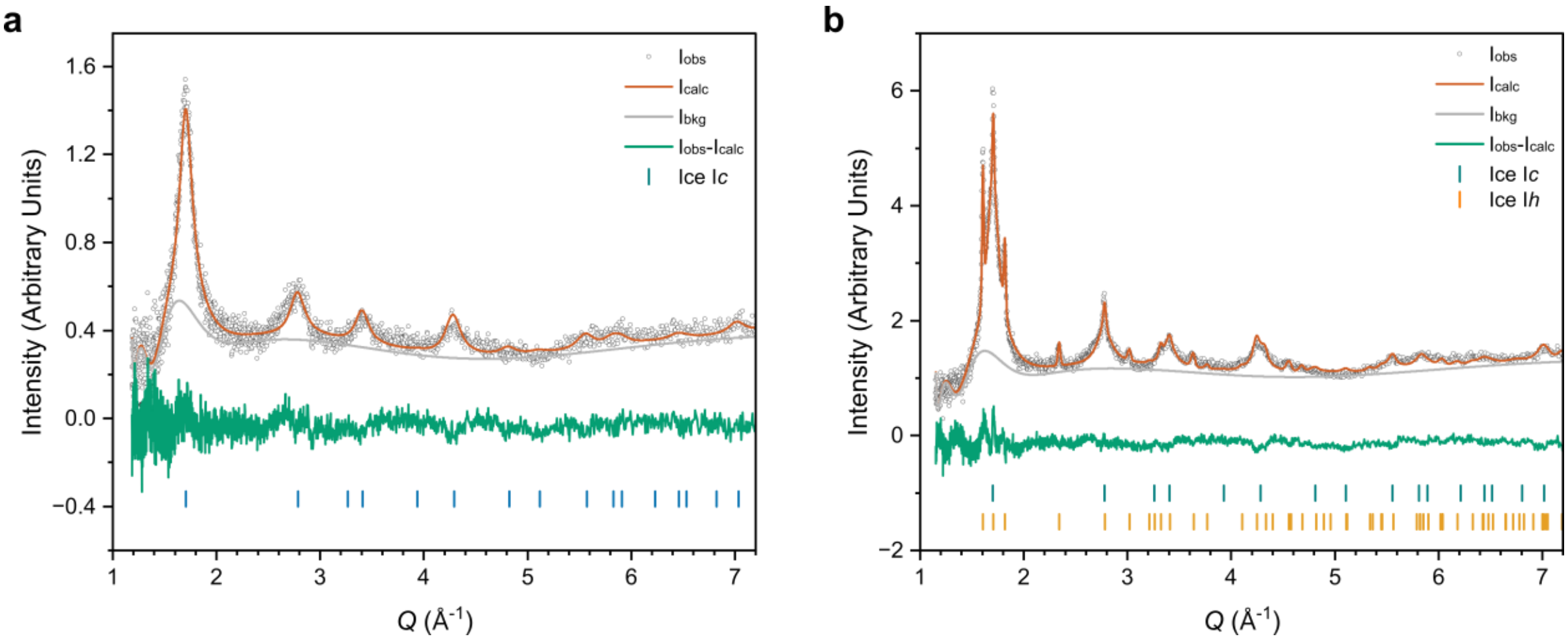

**a.** The diffraction pattern for 1:17 molar ratio $C_4D_8O/D_2O$ at liquid loading 0.34 cm$^3$/g silica (grey points) is well fitted by Rietveld refinement using only cubic ice I*c* (red line). **b**. The diffraction pattern for 1:17 molar ratio $C_4D_8O/D_2O$ at liquid loading 0.70 cm$^3$/g silica requires both cubic ice I*c* and hexagonal ice I*h* components for successful Rietveld refinement. The residuals $I_{obs}$–$I_{calc}$ (green line) and the positions of the ice I*c* (blue ticks) and I*h* (orange ticks) are shown in the panel. In both cases the background $I_{bkg}$ (grey line) accounts for the diffuse scattering from the amorphous fraction within the pores. Refinement on the neutron data was conducted using GSAS-II software. Fitting parameters are summarized in Table 1.

**Extended Data Fig. 5: Controls for Rietveld refinement of 1:17 molar ratio $C_4D_8O/D_2O$ at a liquid loading of 0.60 $cm^3/g$ silica.**

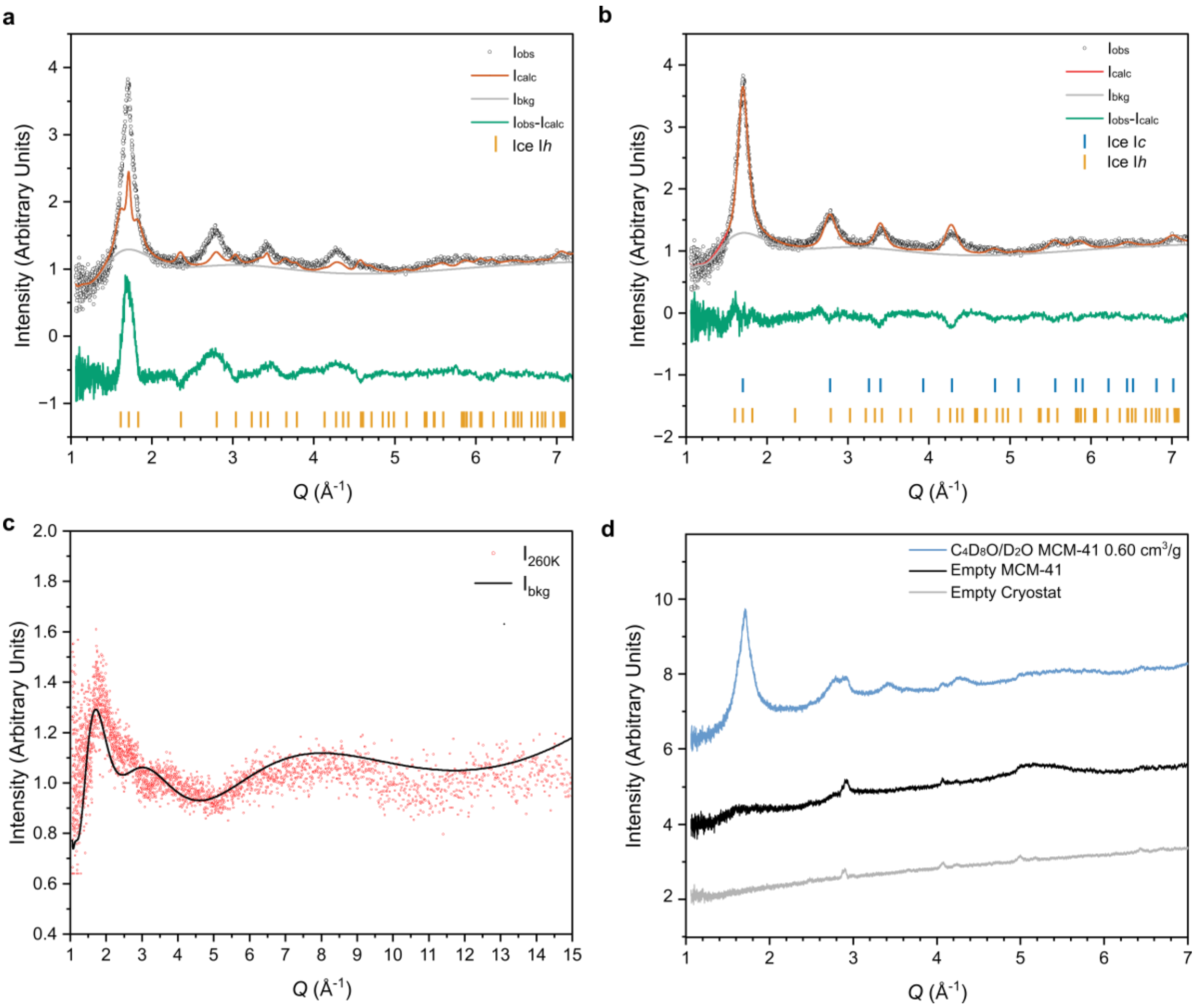

**a.** Best fit based on a single hexagonal ice (I*h*) phase model. **b.** Binary phase model with cubic ice (I*c*) and hexagonal ice (I*h*). The refinement converges with weight fractions w(I*c*) = 0.991 and w(I*h*) = 0.009. **c**. The Rietveld refinement scattering background at a liquid loading of 0.60 $cm^3/g$ silica at 150 K data (black line) compared with an independently measured unfrozen sample at liquid loading of 0.34 $cm^3/g$ silica at 260 K (red dots) under identical normalization. Such comparison shows that the scattering background to the Rietveld refinement is consistent with the presence of an interfacial vitrified THF-water component, with structure akin to that of the confined liquid at lower loading. This interpretation is supported by the strong overlap in the position/intensity of the main diffuse peak at approximately 1.9 $Å^{-1}$ and the overall scattering pattern. Please note that the sample with liquid loading of 0.34 $cm^3/g$ silica at 260 K was used only to check that the Rietveld background is physically reasonable, not in the subtraction itself. **d**. Background neutron diffraction pattern measurements collected by the Polaris neutron diffractometer (ISIS Neutron and Muon Source) at 150 K, of empty cryostat (grey line), liquid-free mesoporous silica (black line), and $C_4D_8O/D_2O$ at liquid loading 0.60 $cm^3/g$ silica, which forms cubic ice I*c* for comparison (blue line). Data offset for clarity.

**Extended Data Fig. 6: Raman evidence for the presence of an interfacial vitrified tetrahydrofuran-water component upon cooling to 150 K.**

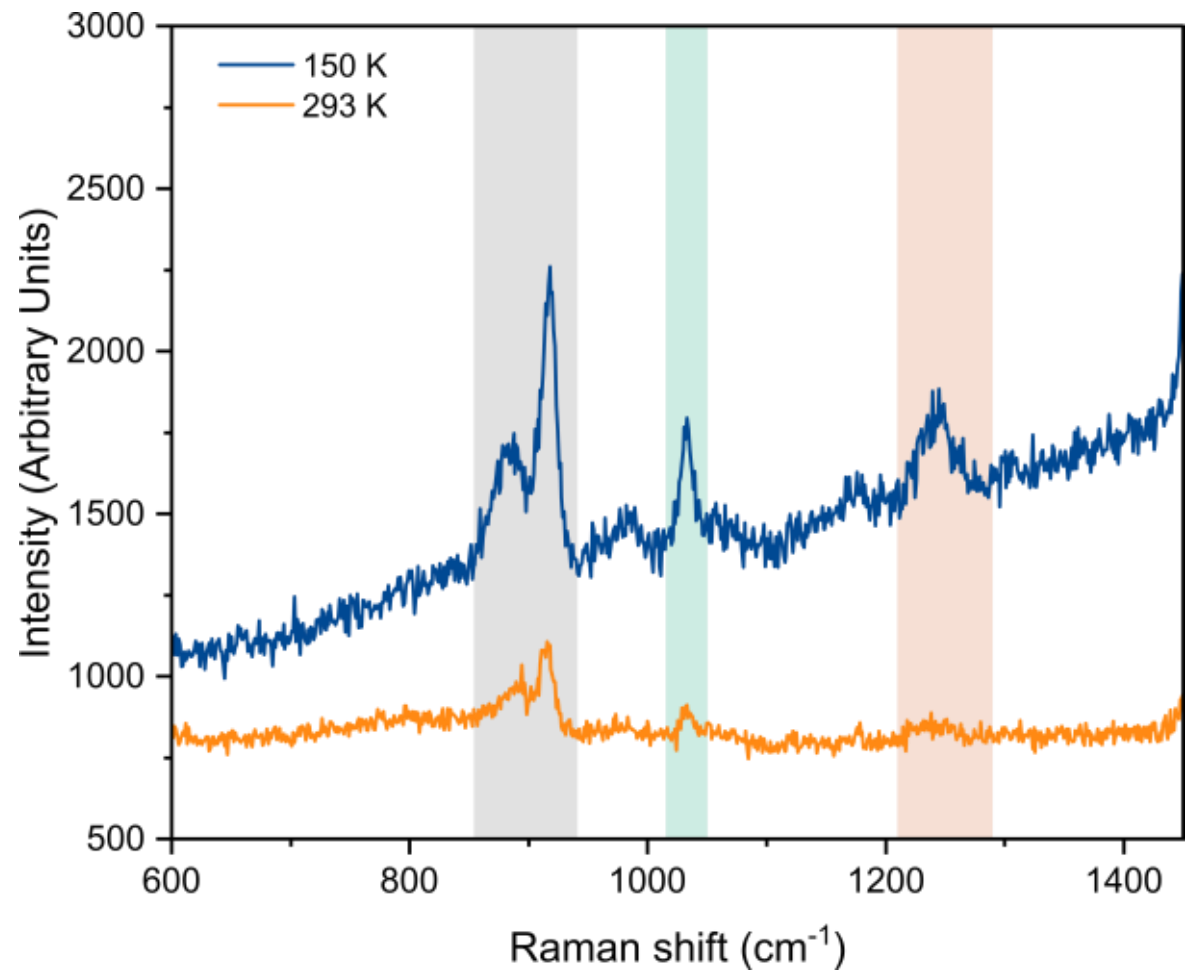

Raman spectra of 1:17 molar ratio $C_4H_8O/D_2O$ at a liquid loading of 0.60 cm$^3$/g silica in MCM-41, cooling from 293 K (orange line) to 150 K (blue line). Bands mark the ring-breathing doublet (grey region 850-940 cm$^{-1}$: 887 cm$^{-1}$ and 918 cm$^{-1}$), the C-O-C stretch (green region 1010-1050 cm$^{-1}$: 1033 cm$^{-1}$), and the C-C/C-O skeletal mode (vermillion region 1210-1290 cm$^{-1}$: 1245 cm$^{-1}$) of THF. On cooling, the band intensities increase sharply, but they neither merge nor shift. It is expected that THF hydrate would show a singlet ring-breathing band at around 922 cm$^{-1}$, and neat crystalline THF at around 915 cm$^{-1}$ [55].

**Extended Data Fig. 7: Raman evidence of tetrahydrofuran locating near the MCM-41 silica surface.**

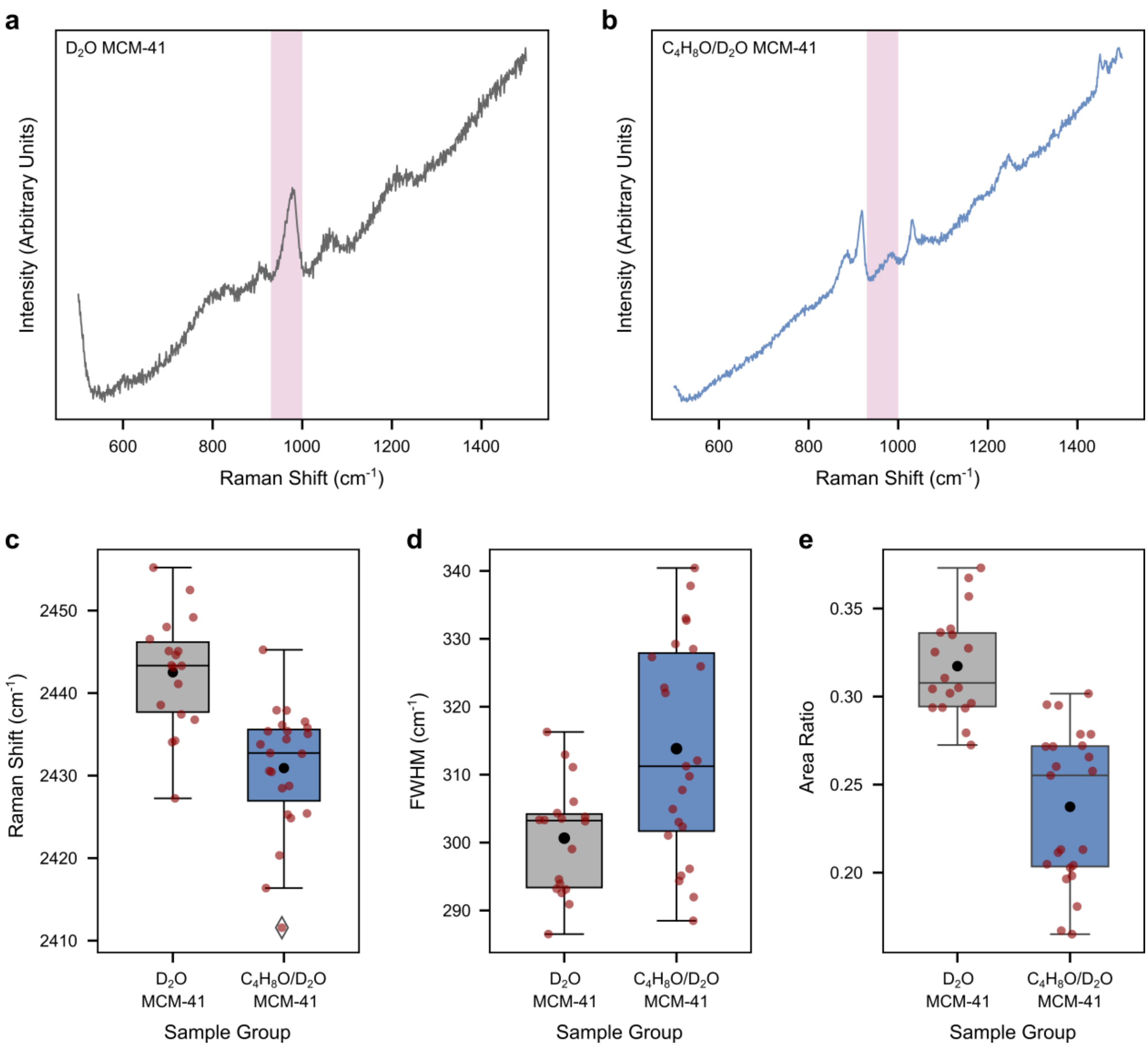

Evidence for tetrahydrofuran molecules locating near MCM-41 pore surface from Raman spectra collected at 293 K and from further analysis on the change of DA (D: HB donor, A: HB acceptor) at 150 K. Pink bands mark the Si-O stretching vibrations of Si-OH groups at MCM-41 pore surface, with **a.** neat $D_2O$ at a liquid loading of 0.60 $cm^3$/g silica and **b**. 1:17 molar ratio $C_4H_8O/D_2O$ at a liquid loading of 0.60 $cm^3$/g silica. A significant decrease in the peak intensity at the presence of THF molecules, indicating the interaction of THF molecules with the silanol groups. **c**. Raman shift of the DA component, which has been suggested to be related to surface-adsorbed water[43]. Neat $D_2O$ at liquid loading 0.60 $cm^3$/g silica in MCM-41 (grey) and 1:17 molar ratio $C_4H_8O/D_2O$ at liquid loading 0.60 $cm^3$/g silica in MCM-41 (blue). Red dots, individual spectral measurements (n = 14 points per group); Horizontal lines, the median values; Black bold dots, mean values. The presence of THF red-shifts the frequency of DA component by ~11.61 $cm^{-1}$ (mean value). Grey diamond represents an outlier of the data, which was not considered for mean and median value calculation. **d.** Full-width at half-maximum (FWHM). The presence of THF broadens the peak width by ~13.2 $cm^{-1}$ (mean value). **e.** Area-ratio of DDAA to the DA components. A clear and sizeable decrease of around 25% is noted in the presence of THF. Relative lower fraction of DDAA water, larger fraction of DA water. **f.** Schematic illustration. The simplest model consistent with these spectral changes is an amorphous 'glassy' THF/water component near pore surface, where THF changes the inter-water bonding from DDAA to DA coordination.

**Extended Data Fig. 8: Elastic fixed-window neutron scattering for the confined tetrahydrofuran-water at a liquid loading of 0.60 cm$^3$/g silica.**

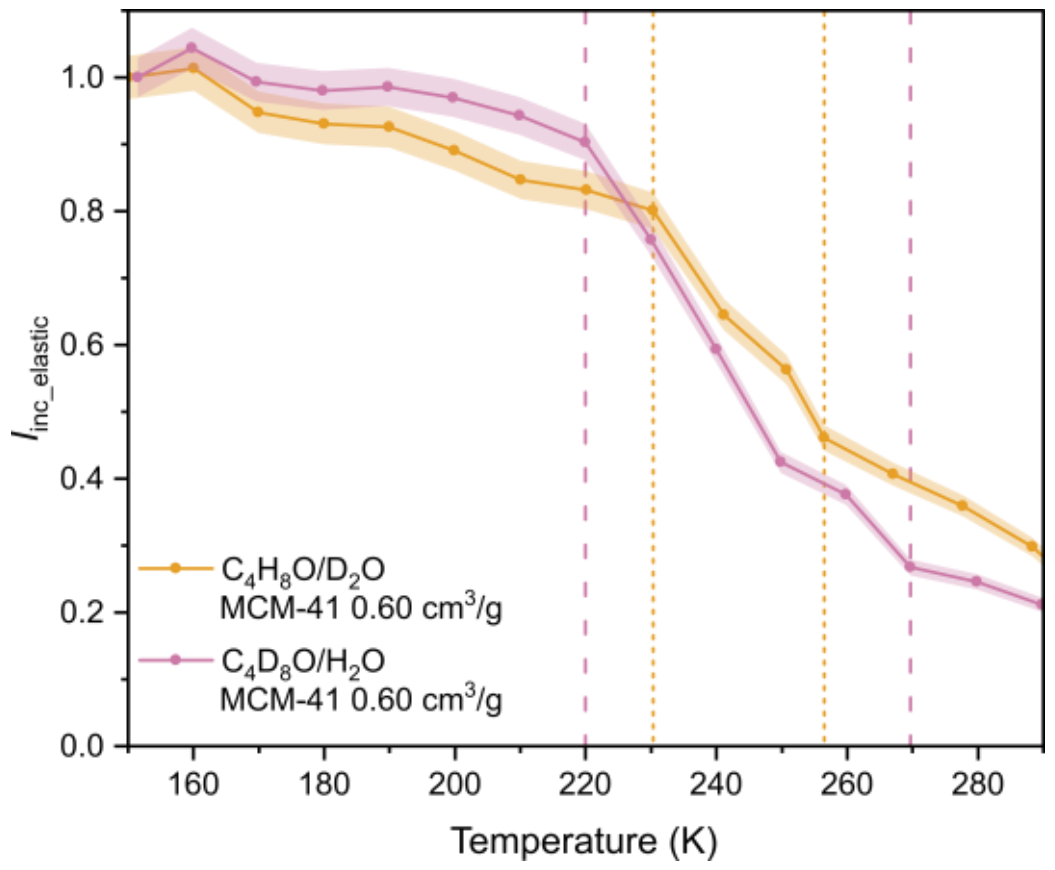

Elastic Fixed Window Scan (EFWS) upon cooling samples from 300 K to 150 K in the OSIRIS beamline. Elastic intensity is integrated over $Q$ = 0.24 - 1.80 Å$^{-1}$ and normalised to 150 K. The signal is dominated by incoherent scattering from protium (H). Selective isotopic substitution of deuterium (D) for H therefore allows resolution of the water and THF dynamics. 1:17 molar ratio $C_4D_8O/H_2O$ at liquid loading 0.60 cm$^3$/g silica (pink line) is a composition with a dynamic scattering signal dominated by water (99.4% of sample incoherent scattering). 1:17 molar ratio $C_4H_8O/D_2O$ at liquid loading 0.60 cm$^3$/g silica (orange line) is a composition with a dynamic scattering signal dominated by THF (90.2% of sample incoherent scattering). Shaded regions represent the uncertainties in the data. Upon increasing temperature, there is a gradual loss of elastic intensity due to the onset of successive molecular motions such as rotation and translation. Inflection points (dashed pink and dotted orange lines) in the EFWS from water (220 K) and THF (230 K) indicate their respective onsets of molecular motion within the temporal window of the spectrometer. Note that the motion of THF molecules is arrested at temperatures higher than the freezing point of bulk THF (167.4K) due to coupling with the interfacial water molecules.

**Extended Data Fig. 9: Quasi-Elastic Neutron Scattering (QENS) determination of the molecular dynamics of the supercooled liquid on the approach to pristine cubic ice I*c*.**

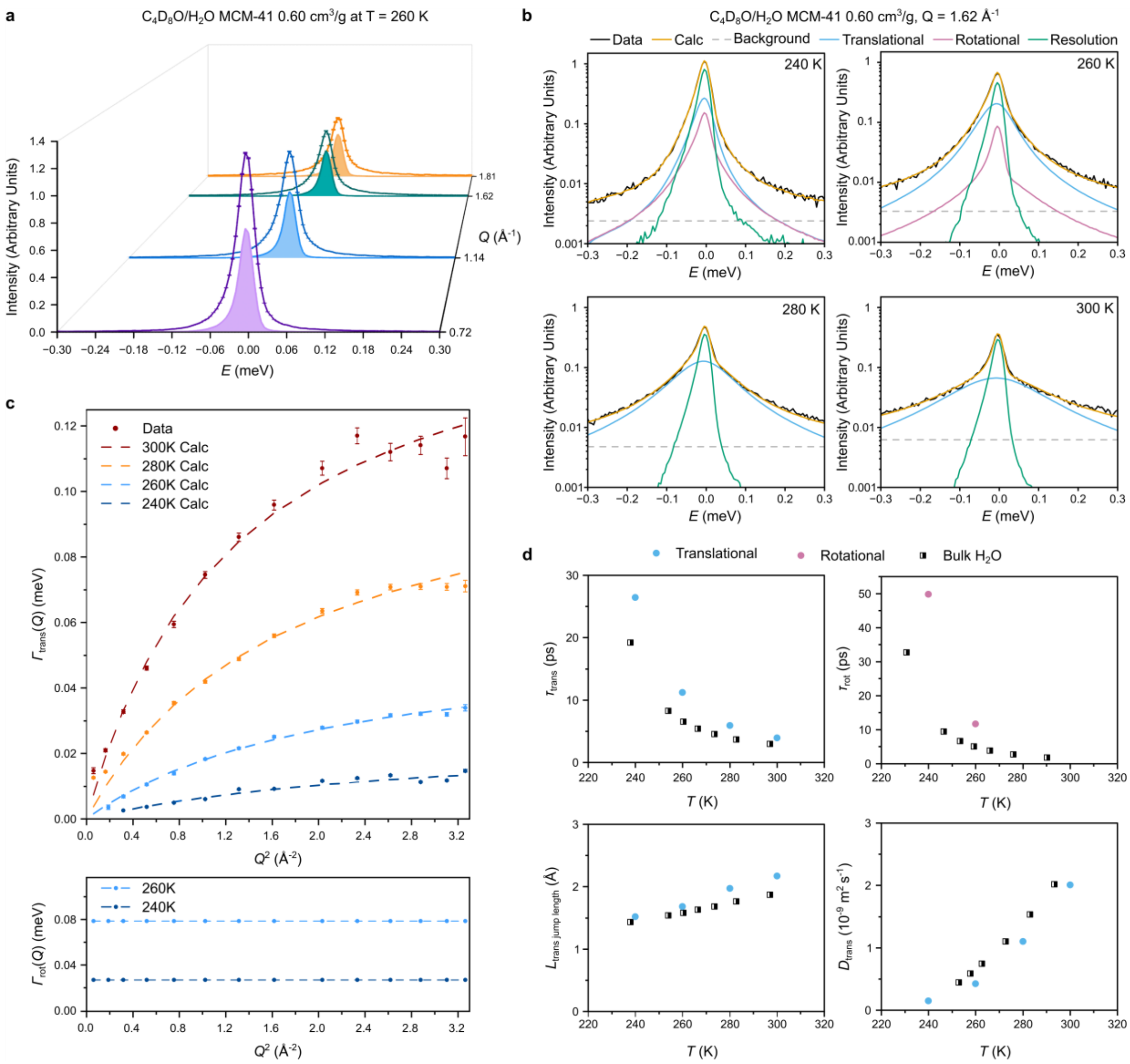

Dynamical insights into the supercooled water from QENS measured at OSIRIS (ISIS), revealing the pathway towards the formation of cubic ice I*c*. **a.** Measured QENS intensity of $C_4D_8O/H_2O$ at liquid loading of 0.60 cm³/g silica at T = 260 K (dots) for selected *Q*-values, compared with the experimental resolution (shaded curve). **b.** QENS data (black line) and fitting across temperature points at 240 K, 260 K, 280 K, and 300 K. Components of the calculated line (orange line) are: a translational term (a *Q*-dependent Lorentzian, blue line), a rotational term (modelled $S(Q,\omega)$ of isotropic H rotations around the centre-of-mass of $H_2O$), pink line, seen only at 240 K and 260 K), a background term (flat, dash grey line), a resolution term (green line) convolved with a delta function. **c.** Half-width at half maximum (HWHM) parameters of the Lorentzian function models (dashed lines) of the translational *Q*-dependent three-dimensional jump diffusion model (top) and the rotational *Q*-independent function (bottom) as a function of temperature. Please note that in this bottom panel, the rotation is fitted to a single Lorentzian across the experimental *Q*-range and hence constrained to be *Q*-independent. **d.** Temperature-dependent analysis of the supercooled $H_2O$ on its translational dynamics (blue points) including residence time ($\tau_{trans}$), jump length *L*, diffusion coefficient ($D_{trans}$) and its rotational dynamics (pink points) including rotational time ($\tau_{rot}$), all compared to bulk water behavior[56].